\documentclass[letters,useAMS,usenatbib]{mnras}
\usepackage{journals}
\usepackage{graphicx}
\usepackage{color}
\usepackage{times}
\usepackage{hyperref}
\usepackage{amsmath}
\usepackage{bm}
\usepackage{multirow}
\usepackage[warn]{textcomp}
\usepackage{rotating}
\usepackage{caption}
\title[SNe~Ia in edge-on galaxies]{Constraining Type Ia supernovae through their heights in edge-on galaxies}
\author[L.~V.~Barkhudaryan]{Lilit~V.~Barkhudaryan\thanks{\selectfont{E-mail:
\href{mailto:l.barkhudaryan@yerphi.am}{l.barkhudaryan@yerphi.am}}}
\\
Center for Cosmology and Astrophysics, Alikhanian National Science Laboratory, 2 Alikhanian Brothers Str., 0036 Yerevan, Armenia}
\begin{document}

\date{Accepted 2022 December 13. Received 2022 November 22; in original form 2022 September 2}

\pagerange{\pageref{firstpage}--\pageref{lastpage}} \pubyear{2022}

\maketitle

\label{firstpage}

\begin{abstract}
  In this Letter, using classified 197 supernovae (SNe) Ia, we perform an analyses of their
  height distributions from the disc in edge-on spirals and investigate their light-curve (LC) decline rates $(\Delta m_{15})$.
  We demonstrate, for the first time, that 91T- and 91bg-like subclasses of SNe~Ia
  are distributed differently toward the
  plane of their host disc. The average height from the disc and its comparison with scales of thin/thick disc
  components gives a possibility to roughly estimate the SNe~Ia progenitor ages:
  91T-like events, being at the smallest heights,
  originate from relatively younger progenitors with ages of about several 100~Myr,
  91bg-like SNe, having the highest distribution, arise from progenitors with significantly older ages $\sim 10$~Gyr,
  and normal SNe~Ia, which distributed between those of the two others, are from progenitors of about one up to $\sim 10$~Gyr.
  We find a correlation between LC decline rates and SN~Ia heights,
  which is explained by the vertical age gradient of stellar population in discs and
  a sub-Chandrasekhar mass white dwarf explosion models, where the $\Delta m_{15}$ parameter is a
  progenitor age indicator.
\end{abstract}

\begin{keywords}
supernovae: individual: Type Ia -- galaxies: disc -- galaxies: stellar content --
galaxies: structure.
\end{keywords}

\section{Introduction}
\label{intro}
\defcitealias{2017MNRAS.471.1390H}{H17}
\defcitealias{2020MNRAS.499.1424H}{H20}

Type Ia supernovae (SNe~Ia) are known to arise from carbon–oxygen (CO) white dwarfs (WDs)
in interacting close binaries. About one-third of SNe~Ia contain unusual
properties and divided into following main subclasses: 91T-like SNe are ${\sim 0.6~{\rm mag}}$ overluminous
than normal SNe~Ia at the $B$-band maximum; 91bg-like events
are ${\sim 2~{\rm mag}}$ subluminous than normal
ones \citep[e.g.][]{2017hsn..book..317T}.
SNe~Ia luminosities at $B$-band maximum and their light curve (LC) decline rates ($\Delta m_{15}$
- difference between magnitudes at the maximum light and those of 15 days)
are correlated \citep{1993ApJ...413L.105P}:
more luminous SNe~Ia have slower declining LCs.

Many studies demonstrated that the progenitor population age of SN~Ia subclasses
is increasing in the sequence of 91T-, normal, and 91bg-like events
\citep[e.g.][]{2001ApJ...554L.193H,2016MNRAS.460.3529A}.
Theoretically, the progenitor age distribution at the current epoch
should have a bimodal shape, with the first peak being below or close to 1~Gyr
and corresponding to the young/prompt SNe~Ia,
and the second peak being at about several Gyr and including old/delayed events
\citep[e.g.][]{2014MNRAS.445.1898C}.

In \citet[][hereafter \citetalias{2017MNRAS.471.1390H}]{2017MNRAS.471.1390H},
taking into account that the height from the disc plane is an indicator of stellar population age
\citep[e.g.][]{2005AJ....130.1574S, 2006AJ....131..226Y,2018MNRAS.475.1203C},
we showed that the majority of SNe~Ia are localized in the discs of edge-on galaxies
and they have about two times larger scale height than core-collapse (CC) SNe, whose progenitors'
ages are up to $\sim$100 Myr.
Also, we showed that the scale height of SNe~Ia
is compatible with that of the older thick disc population of the Milky Way (MW) galaxy.
Nevertheless, we did not investigate different subclasses of SNe~Ia separately.
In this Letter, for the first time, we attempt to accomplish this by studying
the distributions of heights of various SN~Ia subclasses from the host discs.

Recently, in \citet[][hereafter \citetalias{2020MNRAS.499.1424H}]{2020MNRAS.499.1424H},
we verified an earlier finding on the correlation between
LC decline rates of SNe~Ia and the global ages of their host galaxies:
SNe~Ia from older and younger stellar populations, respectively, have larger and smaller $\Delta m_{15}$ values.
This result can be interpreted within the frameworks of the sub-Chandrasekhar mass
($M_{\rm Ch} \approx 1.4 M_{\odot}$) WD explosion models.
The explosion mechanism is realized in the double detonation of a sub-$M_{\rm Ch}$ WD,
in which accreted helium shell detonation initiates second detonation in the core of CO WD
\citep[e.g.][]{2010ApJ...714L..52S,2017MNRAS.470..157B,2017ApJ...851L..50S}.
More luminous SNe~Ia that have slower declining LCs (smaller $\Delta m_{15}$ values)
are produced by the explosion of more massive sub-$M_{\rm Ch}$ WD,
because the luminosity of SN~Ia is related to the mass of $^{56}$Ni synthesized during the WD explosion
\citep[e.g.][]{2006A&A...450..241S}, which in turn is related to the mass of the WD
\citep[see e.g.][for a variety of specific explosion models]{2014MNRAS.438.3456P,2018ApJ...854...52S}.
On the other hand, more massive WD would come from more massive main-sequences stars,
which have shorter lifetime than the progenitors of less massive WDs.
In addition, due to the gravitational wave emission,
massive WDs in the binary system would interact in a shorter timescale.
Thus, it should follow that the LC decline rate $\Delta m_{15}$ of SN~Ia
is correlated with the age of the SN progenitor system
\citep[e.g.][]{2017ApJ...851L..50S,2021ApJ...909L..18S}.
Given this, in our study we simply check the potential correlation
between the SN~Ia heights from host discs and their LC decline rates,
which may provide an indication that both parameters are appropriate stellar population age indicators.

\section{Sample selection and reduction}
\label{samplered}

In this study, to ensure a sufficient number of SNe and to appropriately measure
the SN heights from their host galactic discs, we selected the spectroscopically
classified SN~Ia subclasses (normal, 91T- and 91bg-like)
with distances ${\leq 200~{\rm Mpc}}$ from the Open Supernova Catalog \citep{2017ApJ...835...64G}.
In order to have high confidence on the SN subclasses,
the information is additionally verified utilizing data from
the Weizmann Interactive Supernova data REPository \citep{2012PASP..124..668Y},
Astronomer's Telegram, website of the Central Bureau for Astronomical Telegrams,
etc.\footnote{The sources provide also the equatorial coordinates of the selected SNe~Ia.}

Since we are interested in SNe~Ia that exploded in highly inclined spiral galaxies,
we need to roughly classify the morphology and estimate the inclination of hosts.
To perform this we employed the Sloan Digital Sky Survey
Data Release 16 \citep[DR16;][]{2020ApJS..249....3A},
the SkyMapper DR2 \citep[][]{2019PASA...36...33O}, and
the Pan-STARRS DR2 \citep[][]{2016arXiv161205560C},
which together cover the whole sky and provide the $gri$ bands composed images for each host galaxy.
Hosts with visible low inclinations $(i \lesssim 60^\circ)$
and obviously elliptical, lenticular, or irregular morphology were excluded from the study.
Following \citetalias{2020MNRAS.499.1424H},
we further morphologically classified the hosts and
created the $25~{\rm mag~arcsec^{-2}}$ elliptical apertures for each galaxy
on the surveys' $g$-band images enabling exact measurements of the SN hosts' inclinations,
semi-major $(R_{25})$ and semi-minor $(Z_{25})$ axes.

The next step was to use the estimated elongations $(R_{25}/Z_{25})$ and morphological types
of galaxies to calculate inclinations, following the approach of \citet{1997A&AS..124..109P}.
It is worth noting that the calculated inclinations for galaxies with prominent bulges are inaccurate,
as the isophotes of bulges in highly inclined galaxies reduce the real galaxy disc inclinations.
Such scenarios got special attention for exact inclination calculation,
with only the isophotes of discs being taken into account
(see \citetalias{2017MNRAS.471.1390H}).
Finally, we limited the sample of host galaxies to those with an inclination of $80^\circ \le i \le 90^\circ$.
As a result, we sampled 196 S0/a--Sdm galaxies with a nearly edge-on view,
where a total of 197 SNe~Ia were discovered (Table~\ref{HOSTandSNIa}).

It is important to test the representativeness of our edge-on SN host sample
compared to a sample of galaxies arbitrarily aligned along line-of-sight.
Using the two-sample Kolmogorov--Smirnov (KS) and Anderson--Darling (AD)
tests,\footnote{We used a 5 per cent significance level
as the threshold for the tests.} we compared the distributions of
the sampled SN~Ia subclasses and the morphological types of their
hosts with the same distributions of nearly complete volume-limited
(${\leq 80~{\rm Mpc}}$) sample of the Lick Observatory Supernova
Search \citep[LOSS;][]{2011MNRAS.412.1441L}. In our and LOSS samples, the
representations of SN~Ia subclasses are not statistically different
(the probabilities that the distributions are drawn from the same
parent sample are $>0.6$). The frequencies of morphologies
in our and LOSS samples are also consistent between each other
(probabilities are $>0.1$). Therefore, any artificial loss or excess of
SN subclasses and/or host's morphologies should be not significant in our sample.

We used the methods described in our earlier study on edge-on SN hosts
(\citetalias{2017MNRAS.471.1390H})
to determine the height $(V)$ of a SN from the plane of its host disc
(i.e. the vertical distance of a SN from the major axis of host),
as well as the projected radius $(U)$ along the plane.
The $g$-band images were used for these measurements.
The values of $U$ and $V$ are given in arcsec units.
In this study, we used the $R_{25}$ normalization (in arcsec) to bring the galaxies
to relatively the same size (normalized height is $V/R_{25}$,
normalized projected radius is $U/R_{25}$).
For more details on the measurement techniques, the reader is referred to
\citet{2016MNRAS.456.2848H}, \citetalias{2017MNRAS.471.1390H}.
In the Appendix, we also applied $Z_{25}$ normalization to the $V$ parameter $(V/Z_{25})$.

Recall that the galaxies in our sample have an inclination of $80^\circ - 90^\circ$,
which can introduce discrepancies in (projected) height measurements when compared to
physical heights in the same galaxies with inclination of $90^\circ$.
To check the impact of this effect, using a Monte Carlo (MC) simulation,
we generated 1000 SN heights in $i=90^\circ$ disc
adopting a generalized vertical distribution
$(f(\tilde{z})={\rm sech}^{2}(\tilde{z}/\tilde{z}_{0})$, where $\tilde{z}=V/R_{25})$
and its scale heights for SNe~Ia in spiral galaxies
($\tilde{z}_{0}=0.083$; \citetalias{2017MNRAS.471.1390H}).
Then, we randomly assigned an inclination within $80^\circ - 90^\circ$
to the disc for each SN and estimated the projected SN height from the major axis of the host.
Eventually, the comparison of the generated and projected heights
showed that the differences between them is not significant
($P_{\rm KS}=0.130$, $P_{\rm AD}=0.200$).
Thus, the mentioned effect has a minor impact on the real height measurements,
which accounts, on average, for about 10 per cent of the measured value
($\pm0.01$ in absolute units, typically within the range of measurement errors).

Because we aim to investigate the possible relationships between photometric features
like $\Delta m_{15}$ and the heights of SNe~Ia from the disc,
following \citetalias{2020MNRAS.499.1424H}, we conducted a thorough literature search to
assemble the $B$-band LC decline rates for our 197 SNe.
Only 69 of the SNe~Ia in our sample have available $\Delta m_{15}$ values.

Table~\ref{databasetab} provides our database of
197 individual SNe~Ia (SN designation, $U$ and $V$, spectroscopic
subclass and $\Delta m_{15}$ with their sources) and
their 196 hosts (galaxy designation,
distance, morphological type, $R_{25}$, and $Z_{25}$).

\section{Results and discussion}
\label{RESults}

\subsection{Directional distributions of SNe~Ia in edge-on spiral hosts}
\label{RESults1}

Following \citet{2021MNRAS.505L..52H},
for the SNe~Ia in edge-on spirals of the current study,
we perform the two-sample KS and AD tests
comparing the $|V|/R_{25}$ and $|U|/R_{25}$ distributions between each other.
Table~\ref{UVPKSPAD} shows that the bulk of SNe~Ia in all of the SN subclasses
are localized in the host galaxies' discs.
For 91bg-like SNe only, the AD test shows barely significance, unlike the KS test,
which is probably due to the statistics with the smallest sample size.

\begin{table}
  \centering
  \begin{minipage}{84mm}
  \caption{Comparison of the positional distributions of the
           SN~Ia subclasses along major $U$ and minor $V$ axes.}
  \tabcolsep 4pt
  \label{UVPKSPAD}
    \begin{tabular}{lccccrr}
    \hline
  \multicolumn{1}{l}{SN} & \multicolumn{1}{c}{\emph{$N_{\rm SN}$}} & \multicolumn{1}{c}{$\langle|U|/R_{25}\rangle$} & \multicolumn{1}{c}{versus} & \multicolumn{1}{c}{$\langle|V|/R_{25}\rangle$} &  \multicolumn{1}{c}{$P_{\rm KS}^{\rm MC}$} & \multicolumn{1}{c}{$P_{\rm AD}^{\rm MC}$} \\
  \hline
    Normal & 144 & $0.28^{+0.05}_{-0.04}$ & versus & $0.07^{+0.01}_{-0.01}$ & $<$\textbf{0.001} & $<$\textbf{0.001}\\
    91T & 30 & $0.25^{+0.12}_{-0.07}$ & versus & $0.05^{+0.03}_{-0.02}$ & $<$\textbf{0.001} & $<$\textbf{0.001} \\
    91bg & 23 & $0.25^{+0.15}_{-0.08}$ & versus& $0.14^{+0.08}_{-0.04}$ & \textbf{0.022} & 0.145 \\
    All & 197 & $0.27^{+0.04}_{-0.03}$ & versus & $0.07^{+0.01}_{-0.01}$ & $<$\textbf{0.001} & $<$\textbf{0.001}\\
  \hline
  \end{tabular}
  \parbox{\hsize}{\emph{Notes.} The $P_{\rm KS}$ and $P_{\rm AD}$ probabilities that the distributions
                  are drawn from the same parent sample are calculated using a MC
                  simulation with $10^5$ iterations.
                  Each subsample's mean values with their 95 per cent confidence intervals (CIs) are presented.
                  The $P$-values are bolded when differences between the distributions are
                  statistically significant $(P\leq0.05)$.}
  \end{minipage}
\end{table}
\begin{table}
  \centering
  \begin{minipage}{84mm}
  \caption{Comparison of the $|U|/R_{25}$ and $|V|/R_{25}$ distributions between different subclasses of SNe~Ia.}
  \tabcolsep 1.9pt
  \label{UR25VR25}
    \begin{tabular}{lcclccc}
    \hline
  Subsample~1 & \multicolumn{1}{c}{$N_{\rm SN}$} & versus & \multicolumn{1}{c}{Subsample~2} & \multicolumn{1}{c}{$N_{\rm SN}$} & \multicolumn{1}{c}{$P_{\rm KS}^{\rm MC}$}&\multicolumn{1}{c}{$P_{\rm AD}^{\rm MC}$}\\
  \hline
    $|U|/R_{25}$ of Normal & 144 & versus & $|U|/R_{25}$ of 91bg & 23 &  0.279 & 0.166\\
    $|U|/R_{25}$ of Normal & 144 & versus & $|U|/R_{25}$ of 91T & 30 &  0.828 & 0.835\\
    $|U|/R_{25}$ of 91bg & 23 & versus & $|U|/R_{25}$ of 91T & 30 &  0.756 & 0.611\\
    $|V|/R_{25}$ of Normal & 144 & versus & $|V|/R_{25}$ of 91bg & 23 &  0.079 & \textbf{0.010}\\
    $|V|/R_{25}$ of Normal & 144 & versus & $|V|/R_{25}$ of 91T & 30 &  0.685 & 0.588\\
    $|V|/R_{25}$ of 91bg & 23 & versus & $|V|/R_{25}$ of 91T & 30 &  \textbf{0.033} & \textbf{0.022}\\
  \hline
  \end{tabular}
  \small
  \parbox{\hsize}{\emph{Notes.} The explanations for the $P$-values are identical
                  to those in Table~\ref{UVPKSPAD}.}
  \end{minipage}
\end{table}

\begin{figure}
\begin{center}$
\begin{array}{@{\hspace{0mm}}c@{\hspace{0mm}}}
\includegraphics[width=\hsize]{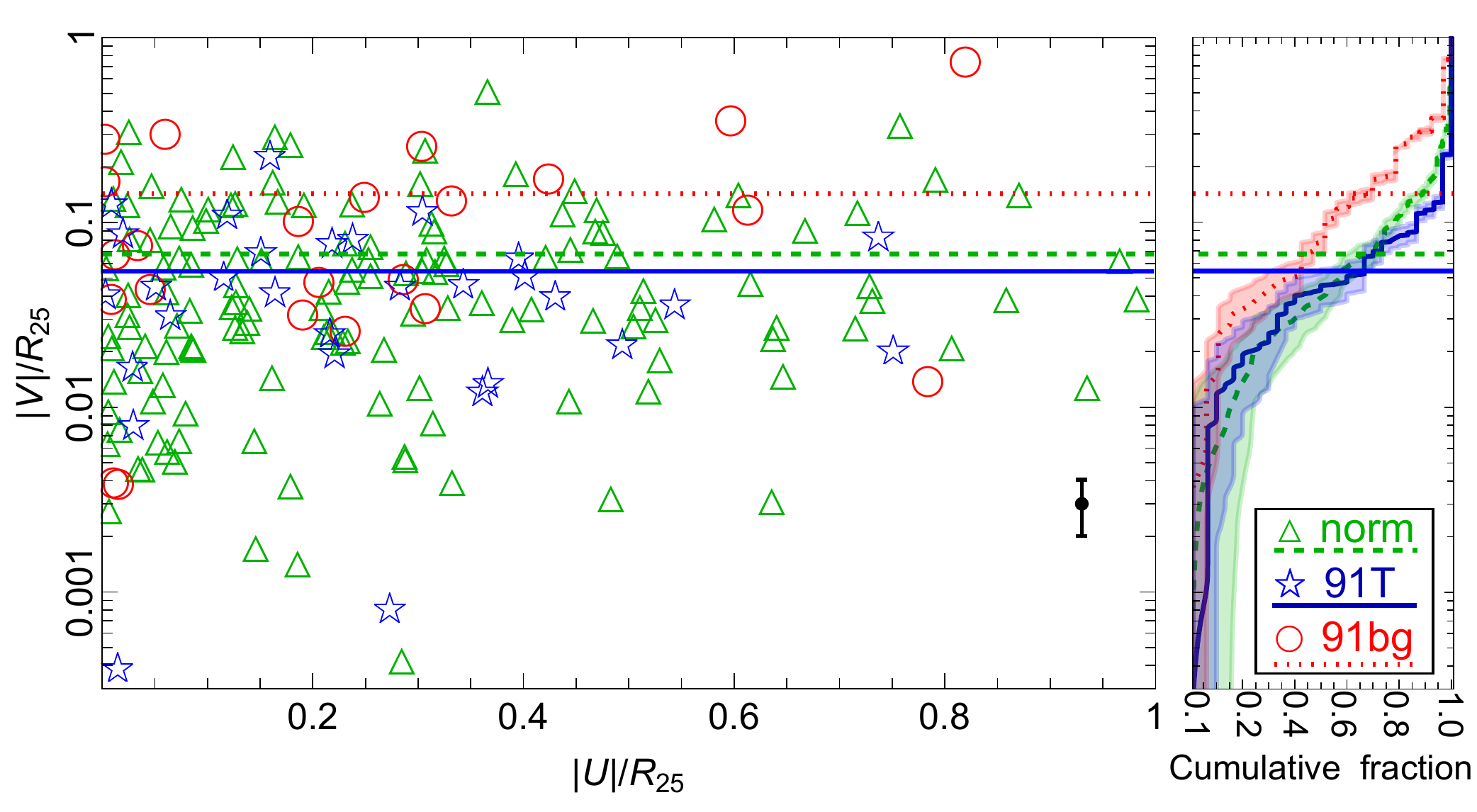}
\end{array}$
\end{center}
\caption{Left panel: distributions of $|V|/R_{25}$ versus $|U|/R_{25}$
        for normal, 91T-, and 91bg-like SNe. The error bar on the right side of the panel shows the
        characteristic error in the height estimation
        due to possible inclination floating in $80^\circ - 90^\circ$.
        The lines show the mean $|V|/R_{25}$
        values for each SN~Ia subclass. Right panel: the heights' cumulative distributions for
        different SNe~Ia.
        The light coloured regions around each curve represent the appropriate spreads
        considering the uncertainties in height measurements.}
\label{VR25UR25VR25CDF}
\end{figure}

We then compare the projected and normalized radii $|U|/R_{25}$ and
the heights $|V|/R_{25}$ between different SN~Ia subclasses.
In Table~\ref{UR25VR25}, the KS and AD tests
show that the radial distributions of normal, 91T- and 91bg-like SNe
are consistent with one another.
In addition, the height distributions of normal
and 91T-like SNe are consistent between each other.
At the same time, the height distributions of 91T- and 91bg-like SNe
are significantly different. The same is happens for the distributions of
normal and 91bg-like SNe (with barely KS test significance).
Fig.~\ref{VR25UR25VR25CDF} shows a scatterplot of $|V|/R_{25}$ versus $|U|/R_{25}$,
and the cumulative distributions of $|V|/R_{25}$ values for different SN~Ia subclasses.
The 91T-like SNe have the smallest height distributions, closest to the disc plane,
whereas the 91bg-like SNe have the highest distribution
(see also the $\langle|V|/R_{25}\rangle$ values in Table~\ref{UVPKSPAD}).
Normal SNe~Ia have a height distribution that is somewhat between those of the two others,
but closer to 91T-like events.

The results, in Table~\ref{UVPKSPAD}, are in agreement with those of our previous papers
(\citetalias{2017MNRAS.471.1390H}, \citealt{2016MNRAS.456.2848H,2021MNRAS.505L..52H}).
As already stated, the disc, rather than the spherical component,
is where the majority of normal, 91T- and 91bg-like SNe~Ia in spiral galaxies arise.
In particular, \citet{2014MNRAS.445.1898C} showed that SN~Ia progenitor age distribution
in spirals peaks up to several hundred Myr ($<1$~Gyr) and has a long tail up to $\sim10$~Gyr,
implying that the bulk of progenitors come from young/intermediate
stellar component \citep[e.g.][]{2012PASA...29..447M},
which is mostly found in discs \citep[e.g.][]{2018A&A...614A..48B}.
Although the old stars ($>1$~Gyr) are distributed in both disc and
spherical components of spiral galaxies, most of the old progenitors of SNe~Ia
also distributed in discs, as evidenced by the behaviour of
the bulge to disc (B/D) flux ratio \citep[e.g.][]{2008MNRAS.388.1708G}
or mass ratio \citep[e.g.][]{2018A&A...614A..48B}.
For example, the B/D flux ratio in the $K$-band,
which is a tracer of the old population,
decreases passing from early- to late-type spirals \citep[e.g.][]{2008MNRAS.388.1708G}:
the average B/D flux ratios are $\sim{0.33}$ and $\sim{0.07}$ for
S0/a--Sbc and Sc--Sdm galaxies, respectively.

Now let us look at the results in Table~\ref{UR25VR25}.
Recall that we only use edge-on host galaxies, therefore it is practically impossible to correct
the discs for the inclination effect and properly study the de-projected radial distributions of SNe.
This could explain why the projected radial distributions of normal, 91bg, and 91T-like SNe are all consistent
(see Table~\ref{UR25VR25}),
although it is known that the ages and other parameters of various stellar populations
in spiral galaxies demonstrate radial dependency \citep[e.g.][]{2015A&A...581A.103G}.

Table~\ref{UR25VR25} also shows, statistically, that 91T- and 91bg-like SNe~Ia
are distributed differently toward the plane of their host disc.
The mean heights are growing, starting with 91T-like events and
progressing through normal and 91bg-like SNe (Table~\ref{UVPKSPAD}).
On the other hand, it is well-known that spiral galaxies have a vertical stellar age gradient,
with the age increasing as the vertical distance from the disc plane increases
\citep[e.g.][]{2005AJ....130.1574S, 2006AJ....131..226Y,2018MNRAS.475.1203C}.
Therefore, from the perspective of the vertical distribution (an age tracer)
it may be deduced that the progenitors of 91T-like and normal SNe~Ia are relatively younger
than those of 91bg-like events.
At least the age differences should be significant for
91T- versus 91bg-like SNe (Table~\ref{UR25VR25}, Fig.~\ref{VR25UR25VR25CDF}).
The results are unaffected when the $Z_{25}$ normalization is applied (Table~\ref{VVR25}).
We emphasize that the current study is the first to demonstrate the observational differences
in the heights of the SN~Ia subclasses.

In fact, more luminous 91T-like SNe could be found more easily at the brighter host
galaxy background than less luminous 91bg-like events.
This would mean that 91T-like SNe could be observed closer to the disc than 91bg-likes.
If so, the observed effect would be a selection bias.
However, it is crucial to note that 91T-like SNe are not as frequently detected
at higher heights as 91bg-likes (see Fig.~\ref{VR25UR25VR25CDF}).
More luminous 91T-like SNe would undoubtedly be found if they had exploded at
the higher heights from the disc.
Hence, it is likely that the detection of 91T-like SNe at lower heights as opposed
to 91bg-likes is a real effect rather than the product of the mentioned selection bias.
This is further supported by the observation that 91T-like SNe are mostly associated
with star-forming environments
(e.g. \citealt{2009ApJ...707...74R}; \citealt{2013MNRAS.429.1425R}; \citetalias{2020MNRAS.499.1424H})
than 91bg-like SNe, which are more frequently seen in older environments
\citep[e.g.][]{2019PASA...36...31P}.
On the other hand, the star-forming environment has the lowest height in the galactic disc
\citep[e.g.][]{2008ApJ...673..864J}.

\subsection{Constraining the age of SN~Ia progenitors}
\label{RESults2}

It is noteworthy that along with the qualitative age constraints of SN~Ia progenitors
we can add also quantitative ones.
Table~\ref{ThickThindisc} compares the scale heights of SN~Ia subclasses in our sample
with the exponential scale heights of the MW thin and thick discs,
as well as with those of 141 edge-on S0/a--Sd galaxies from \citet{2018AA...610A...5C},
sampled according to the different morphological groups.
The scale height of CC SNe
in late-type host galaxies is also shown from our previous paper \citetalias{2017MNRAS.471.1390H}.
Here an exponential vertical distribution $\exp(-|\tilde{z}|/\widetilde{H})$
is used, where the scale height $\widetilde{H}$ is normalized to the galaxy $R_{25}$ radius.
The scale height of SNe $\widetilde{H}_{\rm SN}=\langle|V|/R_{25}\rangle$
for an exponential vertical distribution
(see \citetalias{2017MNRAS.471.1390H}, for more details).
Because the scale height of a stellar population depends on the morphological type of galaxies,
being larger in early-types \citep[e.g.][]{2006AJ....131..226Y,2014ApJ...787...24B},
we split the sample into early- and late-type hosts in Table~\ref{ThickThindisc}
to accurately compare different scales.
Note that in spiral galaxies the majority of 91T-like events are found in Sb--Sdm (late-type) morphological bin,
while most of normal SNe~Ia and 91bg-like events are distributed in S0/a--Sc (early-type) bin
(Table~\ref{HOSTandSNIa}, see also \citetalias{2020MNRAS.499.1424H}).

As shown in Table~\ref{ThickThindisc}, in early-type spirals,
the scale height of normal SNe~Ia is found between those of the thick and thin discs,
while the scale height of 91bg-like events is clearly consistent with the thick disc.
In late-type spirals, the scale height of 91T-like SNe~Ia is close to that of CC SNe, while being larger.
The average height of normal SNe~Ia again is between thin and thick discs.
The scale height of 91bg-like events again is in agreement with those of thick discs.
This is a rough comparison when taking into account the error bars of the mean heights,
it nevertheless gives us a numerical understanding of the relative vertical distributions of SN subclasses
in comparison with thin and thick components of galactic disc.

\begin{figure}
\begin{center}$
\begin{array}{@{\hspace{0mm}}c@{\hspace{0mm}}}
\includegraphics[width=\hsize]{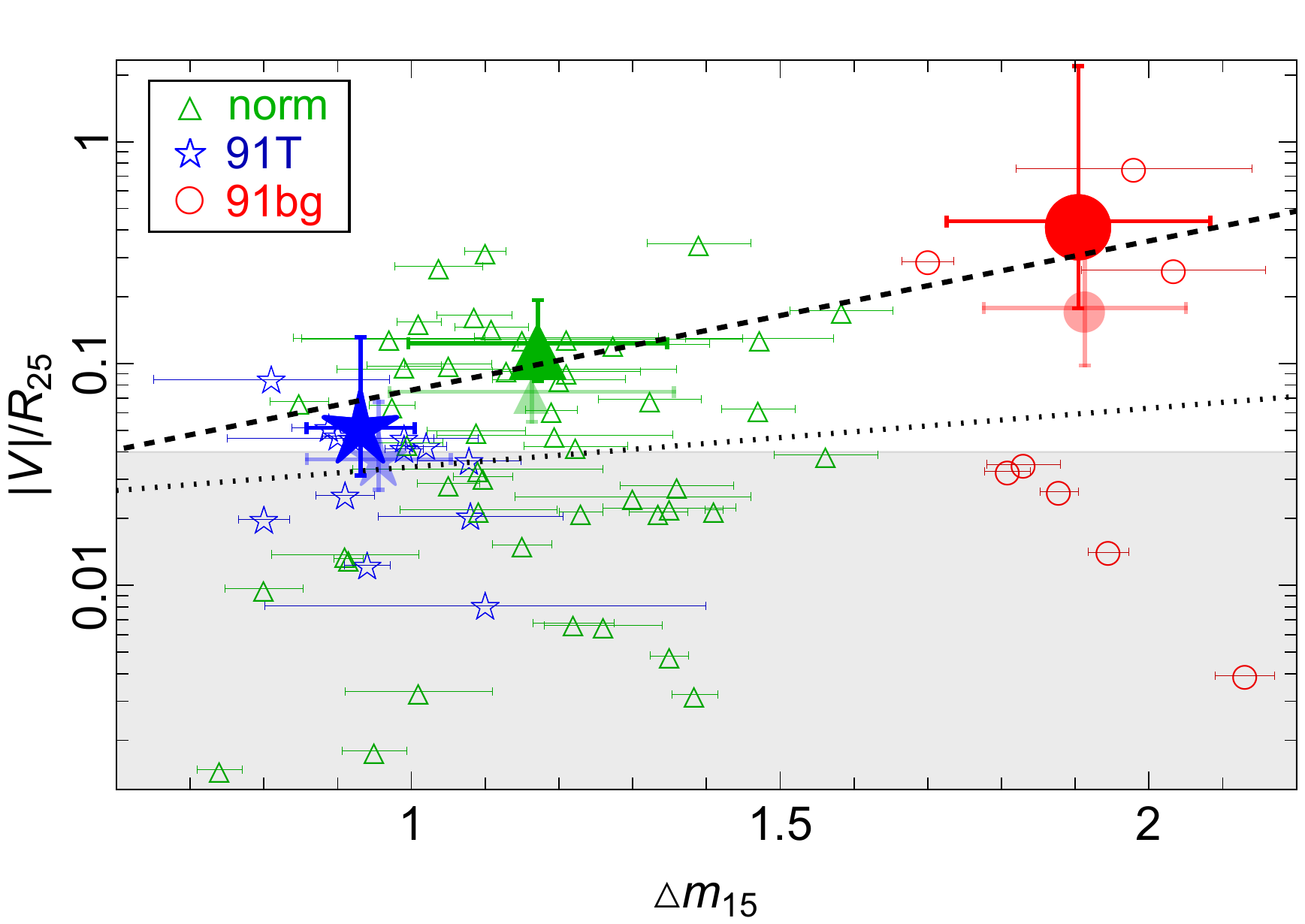}
\end{array}$
\end{center}
\caption{Distributions of $|V|/R_{25}$ versus $\Delta m_{15}$ for different SN~Ia subclasses.
        The dotted and dashed lines, which encompass all SN Ia subclasses,
        present the best-fitting lines for entire and dust-truncated (outside the shaded area)
        discs, respectively.
        Averaged values of $|V|/R_{25}$ (and $\Delta m_{15}$) with their
        95 per cent CIs (and $1\sigma$ errors)
        for entire and dust-truncated samples are presented by
        medium-transparent and big-filled symbols, respectively.}
\label{VR25m15}
\end{figure}

On the other hand,
\citet{2006AJ....131..226Y} found that the scale height of thin disc of 34 late-type
spiral galaxies corresponds to those of young/intermediate stellar populations
with ages from $\sim 10$~Myr up to a few Gyr,
while the scale height of thick disc is comparable to those of old stellar population
with ages from a few Gyr up to $\sim 10$~Gyr \citep[see also][]{2005AJ....130.1574S,2008ApJ...683..707Y}.
A similar result was obtained by
\citet{2021A&A...645L..13C} for the thick disc,
where the age of stellar population increases from $\sim 5$ to $\sim 10$~Gyr
\citep[see also][]{2015A&A...584A..34C,2018AA...610A...5C,2016MNRAS.460L..89K}.
Similarly, the ages of stellar populations of the MW thin and thick discs are estimated to be
up to a few Gyr and from a few Gyr up to $\sim 10$~Gyr, respectively
\citep[e.g.][]{2007ApJ...663L..13B,2008ApJ...673..864J}.
Notably, CC SNe arise from young progenitors with ages up to $\sim100$~Myr and their vertical extend
is accordingly less than the thin disc (\citetalias{2017MNRAS.471.1390H}).

Thus, the various SN~Ia subclasses correspond to different stellar population ages
being distributed at the various average heights from the disc
(e.g. \citealt{2005AJ....130.1574S}; \citealt{2008ApJ...683..707Y};
\citetalias{2017MNRAS.471.1390H}).
From Table~\ref{ThickThindisc}, we can impose rough numerical constraints on the SN progenitors:
91T-like events arise from progenitors with ages about several 100~Myr,
the ages of progenitors of 91bg-like SNe are comparable to $\sim 10$~Gyr,
while normal SNe~Ia arise from progenitors with ages from about one up to $\sim 10$~Gyr.

It is important to note that the delay
time\footnote{Time interval between the progenitor formation and the SN explosion.}
of the SN progenitor system
could be dominated by the timescale of the gravitational inspiral of WDs
in comparison with the stellar age
\citep[a lifetime till it becomes WD, see][]{2013FrPhy...8..116H,2018PhR...736....1L}.
In this study, however, we consider the stellar age of SN progenitors rather
than the system's entire delay time when comparing their average heights with
those of various disc components.
Note that a significant change in the mean vertical scales of the young and old stellar
populations is not expected during the SN progenitor stellar age or the delay time of the systems.
As mentioned above, 91T-like events (and the most of normal~SNe Ia) are
associated with star-forming environments
\citep[$\lesssim500$ Myr;][]{2009ApJ...707...74R,2013MNRAS.429.1425R},
therefore the effect of the gravitational inspiral's timescale should play
a role mostly for 91bg-like SNe.

\subsection{Relating LC decline rates with SN heights from host disc}
\label{RESults3}

SNe~Ia span a variety of properties from subluminous SNe with fast-declining LCs
to overluminous and slowly evolving events \citep[e.g.][]{2017hsn..book..317T}.
The majority of earlier theoretical studies have failed to fit the full range of
observed SNe~Ia properties with a single explosion/progenitor scenario
\citep[see reviews by][]{2013FrPhy...8..116H,2018PhR...736....1L}.
Fortunately, recent theoretical studies in the sub-$M_{\rm Ch}$ WD
explosion models showed an excellent quantitative agreement with observed photometrical behaviours
of SNe~Ia in the entire range of the \citeauthor{1993ApJ...413L.105P} relation
\citep[e.g.][]{2017MNRAS.470..157B,2017ApJ...851L..50S,2021ApJ...909L..18S}.
As mentioned in the Introduction,
the explosion is realized in the double detonation of a sub-$M_{\rm Ch}$ WD,
where the LC decline rate $\Delta m_{15}$ of SN~Ia
is positively correlated with the age of the SN progenitor system
\citep[e.g.][]{2017ApJ...851L..50S,2021ApJ...909L..18S}.

Numerous researches extensively studied the links between the SNe~Ia LC decline rates and
the global age (or age tracers) of host galaxies, as well as local age at SN explosion sites
(e.g. \citealt{2009ApJ...691..661H}; \citealt{2011ApJ...740...92G};
\citealt{2014MNRAS.438.1391P}; \citealt{2016MNRAS.460.3529A};
\citealt{2018A&A...615A..68R}; \citetalias{2020MNRAS.499.1424H}).
These studies demonstrated that, at different levels of significance, the LC decline rate is
correlated with the global/local age:
the $B$-band $\Delta m_{15}$ values increase with stellar population age.
However, the correlation between SNe~Ia decline rate and the height from the host disc,
which is a reliable age indicator of stellar population, has not yet been investigated.
Here, we intend to fill this gap.

\begin{table}
  \centering
  \begin{minipage}{84mm}
  \caption{The correlation test for the $|V|/R_{25}$ versus $\Delta m_{15}$ parameters.}
  \tabcolsep 5.8pt
  \label{VR25dm15}
    \begin{tabular}{lcccccc}
    \hline
  SN & \multicolumn{1}{c}{$N_{\rm SN}$} & \multicolumn{1}{c}{$\langle|V|/R_{25}\rangle$} & versus & \multicolumn{1}{c}{$\langle\Delta m_{15}\rangle$} & \multicolumn{1}{c}{$r_{\rm s}$} & \multicolumn{1}{c}{$P_{\rm s}^{\rm MC}$} \\
  \hline
    All & 69 & $0.08^{+0.02}_{-0.02}$ & versus & 1.21$\pm$0.32 &  0.118 & 0.334 \\
    All$\dag$ & 36 & $0.14^{+0.06}_{-0.04}$ & versus & 1.18$\pm$0.29 &  0.471 & \textbf{0.004} \\
  \hline
  \end{tabular}
  \small
  \parbox{\hsize}{\emph{Notes.} A coefficient of Spearman's rank correlation $(r_{\rm s}\in[-1;1])$ is a metric for determining
                  how closely two variables are related by a monotonic function.
                  The variables are not independent when $P \leq 0.05$ (highlighted in bold).
                  The $P_{\rm s}^{\rm MC}$ values are generated using permutations with $10^5$ MC iterations.
                  The subsample marked with $\dag$ symbol corresponds to SNe with $|V|/R_{25} \geq 0.04$.}
  \end{minipage}
\end{table}

Fig.~\ref{VR25m15} and the Spearman's rank correlation test
in Table~\ref{VR25dm15} show that the trend between $|V|/R_{25}$ and
$\Delta m_{15}$ is positive, but not statistically significant.
At low heights, in Fig.~\ref{VR25m15}, we observe all the SN~Ia subclasses
(full range with slower and faster declining LCs),
but with increasing height, the decline rate of objects increases on average.
However, it should be taken into account that due to the dust extinction in galactic disc
the discovery of SNe~Ia in edge-on galaxies is complicated and biased against objects
at lower heights from the disc \citep[e.g.][]{2015MNRAS.446.3768H}.
The impact of this effect would be greatest on subluminous SNe (91bg-like events).

In late-type galaxies, the vertical distribution of dust has a scale height
that is $\sim 3$ times less than that of thick disc stars \citep[e.g.][]{2007A&A...471..765B}.
While the dust layer is $\sim 1.5$ times thicker in early-type galaxies \citep[e.g.][]{1982AN....303..245H,2014MNRAS.441..869D},
which is closer to our sample of SNe~Ia host galaxies (see Table~\ref{HOSTandSNIa}).
Therefore, to avoid the possible impact of dust we truncate the heights of SNe
with $|V|/R_{25} \geq 0.04$, leaving 36 SNe~Ia in our sample.
For this dust-truncated sample, the Spearman's rank test reveals a significant positive correlation between
the $|V|/R_{25}$ and $\Delta m_{15}$ parameters (Table~\ref{VR25dm15}, Fig.~\ref{VR25m15}).
The results are unaffected when the $Z_{25}$ normalization is applied (Table~\ref{VV25dm15}).
Thus, despite the limited sample size, we demonstrate for the first time a significant correlation
between LC decline rates and SNe~Ia heights, which is consistent with a sub-$M_{\rm Ch}$ WD explosion models
\citep[e.g.][]{2010ApJ...714L..52S,2017MNRAS.470..157B,2017ApJ...851L..50S}
and vertical age gradient of stellar population in discs
\citep[e.g.][]{2006AJ....131..226Y,2018MNRAS.475.1203C}.

It would be important to verify the results in Tables~\ref{UR25VR25} and \ref{VVR25}
while accounting for the selection effects brought by dust extinction.
However, in these tables we compare the SN positions (importantly heights) between the subclasses,
and after the dust-truncation the samples for 91T- and 91bg-like SNe
become, unfortunately, insufficient to perform the statistical tests.

\section{Conclusions}
\label{DISconc}

In this Letter, we analyse the height distributions of SN~Ia subclasses
(normal, 91T- and 91bg-like)
from their host disc plane using spectroscopically classified 197 SNe in edge-on spiral galaxies
with distances ${\leq 200~{\rm Mpc}}$.
In addition, this study is performed to examine potential links between photometric characteristics of SNe~Ia,
like LC decline rates $(\Delta m_{15})$, and SN heights from the disc.

For the first time, we demonstrate that 91T- and 91bg-like subclasses of SNe~Ia
are distributed differently toward the plane of their host edge-on disc.
On average, the SN heights are rising, beginning with 91T-like events and
progressing through normal and 91bg-like SNe~Ia.
Considering that the height from the disc is a stellar population age indicator
and comparing the mean heights of the SN~Ia subclasses with those of thin and thick discs
with known ages, we roughly estimate that 91T-like events originate from relatively younger progenitors
with ages of about several 100~Myr, the ages of progenitors of normal SNe~Ia are from about one up to $\sim 10$~Gyr,
and 91bg-like SNe~Ia arise from progenitors with significantly older ages $\sim 10$~Gyr.
In addition, we show that the SN~Ia LC decline rates
correlate with their heights from the host disc,
after excluding the selection effects brought by dust extinction.
The observed correlation is consistent with the explosion models of a sub-$M_{\rm Ch}$ mass WD
\citep[e.g.][]{2017MNRAS.470..157B,2017ApJ...851L..50S,2021ApJ...909L..18S}
and the vertical age gradient of stellar population in discs
\citep[e.g.][]{2005AJ....130.1574S, 2006AJ....131..226Y,2018MNRAS.475.1203C}.

Fortunately, a far larger spectroscopic and photometric sample of nearby SNe~Ia
will be made available by the ongoing robotic telescope surveys at various locations throughout the globe
(e.g. All Sky Automated Survey for SuperNovae) and by
the forthcoming Vera~C. Rubin Observatory (the Large Synoptic Survey Telescope),
which will allow for statistically more powerful and accurate analysis.

\section*{Acknowledgements}

We appreciate the anonymous referee's suggestions for enhancing our \emph{Letter}.
I thank my PhD supervisor Dr. Artur Hakobyan for his support.
The work was supported by the Science Committee of RA,
in the frames of the research project \textnumero~21T--1C236.

%%%%%%%%%%%%%%%%%%%%%%%%%%%%%%%%%%%%%%%%%%%%%%%%%%

\section*{Data Availability}

The data underlying this study may be found in the article's
supplementary material online (see Table~\ref{databasetab}, for guidance).

%%%%%%%%%%%%%%%%%%%% REFERENCES %%%%%%%%%%%%%%%%%%

\bibliography{VerticalSNIabib}

\begin{thebibliography}{}
\makeatletter
\relax
\def\mn@urlcharsother{\let\do\@makeother \do\$\do\&\do\#\do\^\do\_\do\%\do\~}
\def\mn@doi{\begingroup\mn@urlcharsother \@ifnextchar [ {\mn@doi@}
  {\mn@doi@[]}}
\def\mn@doi@[#1]#2{\def\@tempa{#1}\ifx\@tempa\@empty \href
  {http://dx.doi.org/#2} {doi:#2}\else \href {http://dx.doi.org/#2} {#1}\fi
  \endgroup}
\def\mn@eprint#1#2{\mn@eprint@#1:#2::\@nil}
\def\mn@eprint@arXiv#1{\href {http://arxiv.org/abs/#1} {{\tt arXiv:#1}}}
\def\mn@eprint@dblp#1{\href {http://dblp.uni-trier.de/rec/bibtex/#1.xml}
  {dblp:#1}}
\def\mn@eprint@#1:#2:#3:#4\@nil{\def\@tempa {#1}\def\@tempb {#2}\def\@tempc
  {#3}\ifx \@tempc \@empty \let \@tempc \@tempb \let \@tempb \@tempa \fi \ifx
  \@tempb \@empty \def\@tempb {arXiv}\fi \@ifundefined
  {mn@eprint@\@tempb}{\@tempb:\@tempc}{\expandafter \expandafter \csname
  mn@eprint@\@tempb\endcsname \expandafter{\@tempc}}}

\bibitem[\protect\citeauthoryear{{Ahumada} et~al.,}{{Ahumada}
  et~al.}{2020}]{2020ApJS..249....3A}
{Ahumada} R.,  et~al., 2020, \mn@doi [\apjs] {10.3847/1538-4365/ab929e}, \href
  {https://ui.adsabs.harvard.edu/abs/2020ApJS..249....3A} {249, 3}

\bibitem[\protect\citeauthoryear{{Ashall}, {Mazzali}, {Sasdelli}  \&
  {Prentice}}{{Ashall} et~al.}{2016}]{2016MNRAS.460.3529A}
{Ashall} C.,  {Mazzali} P.,  {Sasdelli} M.,   {Prentice} S.~J.,  2016, \mn@doi
  [\mnras] {10.1093/mnras/stw1214}, \href
  {https://ui.adsabs.harvard.edu/abs/2016MNRAS.460.3529A} {460, 3529}

\bibitem[\protect\citeauthoryear{{Bensby}, {Zenn}, {Oey}  \&
  {Feltzing}}{{Bensby} et~al.}{2007}]{2007ApJ...663L..13B}
{Bensby} T.,  {Zenn} A.~R.,  {Oey} M.~S.,   {Feltzing} S.,  2007, \mn@doi
  [\apjl] {10.1086/519792}, \href
  {https://ui.adsabs.harvard.edu/abs/2007ApJ...663L..13B} {663, L13}

\bibitem[\protect\citeauthoryear{{Bianchi}}{{Bianchi}}{2007}]{2007A&A...471..765B}
{Bianchi} S.,  2007, \mn@doi [\aap] {10.1051/0004-6361:20077649}, \href
  {https://ui.adsabs.harvard.edu/abs/2007A&A...471..765B} {471, 765}

\bibitem[\protect\citeauthoryear{{Bizyaev}, {Kautsch}, {Mosenkov},
  {Reshetnikov}, {Sotnikova}, {Yablokova}  \& {Hillyer}}{{Bizyaev}
  et~al.}{2014}]{2014ApJ...787...24B}
{Bizyaev} D.~V.,  {Kautsch} S.~J.,  {Mosenkov} A.~V.,  {Reshetnikov} V.~P.,
  {Sotnikova} N.~Y.,  {Yablokova} N.~V.,   {Hillyer} R.~W.,  2014, \mn@doi
  [\apj] {10.1088/0004-637X/787/1/24}, \href
  {https://ui.adsabs.harvard.edu/abs/2014ApJ...787...24B} {787, 24}

\bibitem[\protect\citeauthoryear{{Blondin}, {Dessart}, {Hillier}  \&
  {Khokhlov}}{{Blondin} et~al.}{2017}]{2017MNRAS.470..157B}
{Blondin} S.,  {Dessart} L.,  {Hillier} D.~J.,   {Khokhlov} A.~M.,  2017,
  \mn@doi [\mnras] {10.1093/mnras/stw2492}, \href
  {https://ui.adsabs.harvard.edu/abs/2017MNRAS.470..157B} {470, 157}

\bibitem[\protect\citeauthoryear{{Breda} \& {Papaderos}}{{Breda} \&
  {Papaderos}}{2018}]{2018A&A...614A..48B}
{Breda} I.,  {Papaderos} P.,  2018, \mn@doi [\aap]
  {10.1051/0004-6361/201731705}, \href
  {https://ui.adsabs.harvard.edu/abs/2018A&A...614A..48B} {614, A48}

\bibitem[\protect\citeauthoryear{{Chambers} et~al.,}{{Chambers}
  et~al.}{2016}]{2016arXiv161205560C}
{Chambers} K.~C.,  et~al., 2016, preprint, \href
  {https://ui.adsabs.harvard.edu/abs/2016arXiv161205560C} {} (\mn@eprint
  {arXiv} {1612.05560})

\bibitem[\protect\citeauthoryear{{Childress}, {Wolf}  \& {Zahid}}{{Childress}
  et~al.}{2014}]{2014MNRAS.445.1898C}
{Childress} M.~J.,  {Wolf} C.,   {Zahid} H.~J.,  2014, \mn@doi [\mnras]
  {10.1093/mnras/stu1892}, \href
  {https://ui.adsabs.harvard.edu/abs/2014MNRAS.445.1898C} {445, 1898}

\bibitem[\protect\citeauthoryear{{Ciuc{\v{a}}}, {Kawata}, {Lin}, {Casagrande},
  {Seabroke}  \& {Cropper}}{{Ciuc{\v{a}}} et~al.}{2018}]{2018MNRAS.475.1203C}
{Ciuc{\v{a}}} I.,  {Kawata} D.,  {Lin} J.,  {Casagrande} L.,  {Seabroke} G.,
  {Cropper} M.,  2018, \mn@doi [\mnras] {10.1093/mnras/stx3285}, \href
  {https://ui.adsabs.harvard.edu/abs/2018MNRAS.475.1203C} {475, 1203}

\bibitem[\protect\citeauthoryear{{Comer{\'o}n}}{{Comer{\'o}n}}{2021}]{2021A&A...645L..13C}
{Comer{\'o}n} S.,  2021, \mn@doi [\aap] {10.1051/0004-6361/202040175}, \href
  {https://ui.adsabs.harvard.edu/abs/2021A&A...645L..13C} {645, L13}

\bibitem[\protect\citeauthoryear{{Comer{\'o}n}, {Salo}, {Janz}, {Laurikainen}
  \& {Yoachim}}{{Comer{\'o}n} et~al.}{2015}]{2015A&A...584A..34C}
{Comer{\'o}n} S.,  {Salo} H.,  {Janz} J.,  {Laurikainen} E.,   {Yoachim} P.,
  2015, \mn@doi [\aap] {10.1051/0004-6361/201526815}, \href
  {https://ui.adsabs.harvard.edu/abs/2015A&A...584A..34C} {584, A34}

\bibitem[\protect\citeauthoryear{{Comer{\'o}n}, {Salo}  \&
  {Knapen}}{{Comer{\'o}n} et~al.}{2018}]{2018AA...610A...5C}
{Comer{\'o}n} S.,  {Salo} H.,   {Knapen} J.~H.,  2018, \mn@doi [\aap]
  {10.1051/0004-6361/201731415}, \href
  {https://ui.adsabs.harvard.edu/abs/2018AA...610A...5C} {610, A5}

\bibitem[\protect\citeauthoryear{{De Geyter}, {Baes}, {Camps}, {Fritz}, {De
  Looze}, {Hughes}, {Viaene}  \& {Gentile}}{{De Geyter}
  et~al.}{2014}]{2014MNRAS.441..869D}
{De Geyter} G.,  {Baes} M.,  {Camps} P.,  {Fritz} J.,  {De Looze} I.,  {Hughes}
  T.~M.,  {Viaene} S.,   {Gentile} G.,  2014, \mn@doi [\mnras]
  {10.1093/mnras/stu612}, \href
  {https://ui.adsabs.harvard.edu/abs/2014MNRAS.441..869D} {441, 869}

\bibitem[\protect\citeauthoryear{{Gonz{\'a}lez Delgado} et~al.,}{{Gonz{\'a}lez
  Delgado} et~al.}{2015}]{2015A&A...581A.103G}
{Gonz{\'a}lez Delgado} R.~M.,  et~al., 2015, \mn@doi [\aap]
  {10.1051/0004-6361/201525938}, \href
  {https://ui.adsabs.harvard.edu/abs/2015A&A...581A.103G} {581, A103}

\bibitem[\protect\citeauthoryear{{Graham} \& {Worley}}{{Graham} \&
  {Worley}}{2008}]{2008MNRAS.388.1708G}
{Graham} A.~W.,  {Worley} C.~C.,  2008, \mn@doi [\mnras]
  {10.1111/j.1365-2966.2008.13506.x}, \href
  {https://ui.adsabs.harvard.edu/abs/2008MNRAS.388.1708G} {388, 1708}

\bibitem[\protect\citeauthoryear{{Guillochon}, {Parrent}, {Kelley}  \&
  {Margutti}}{{Guillochon} et~al.}{2017}]{2017ApJ...835...64G}
{Guillochon} J.,  {Parrent} J.,  {Kelley} L.~Z.,   {Margutti} R.,  2017,
  \mn@doi [\apj] {10.3847/1538-4357/835/1/64}, \href
  {https://ui.adsabs.harvard.edu/abs/2017ApJ...835...64G} {835, 64}

\bibitem[\protect\citeauthoryear{{Gupta} et~al.,}{{Gupta}
  et~al.}{2011}]{2011ApJ...740...92G}
{Gupta} R.~R.,  et~al., 2011, \mn@doi [\apj] {10.1088/0004-637X/740/2/92},
  \href {https://ui.adsabs.harvard.edu/abs/2011ApJ...740...92G} {740, 92}

\bibitem[\protect\citeauthoryear{{Hacke}, {Schielicke}  \& {Schmidt}}{{Hacke}
  et~al.}{1982}]{1982AN....303..245H}
{Hacke} G.,  {Schielicke} R.,   {Schmidt} K.~H.,  1982, \mn@doi [Astron.
  Nachr.] {10.1002/asna.2103030406}, \href
  {https://ui.adsabs.harvard.edu/abs/1982AN....303..245H} {303, 245}

\bibitem[\protect\citeauthoryear{{Hakobyan} et~al.,}{{Hakobyan}
  et~al.}{2016}]{2016MNRAS.456.2848H}
{Hakobyan} A.~A.,  et~al., 2016, \mn@doi [\mnras] {10.1093/mnras/stv2853},
  \href {https://ui.adsabs.harvard.edu/abs/2016MNRAS.456.2848H} {456, 2848}

\bibitem[\protect\citeauthoryear{{Hakobyan} et~al.,}{{Hakobyan}
  et~al.}{2017}]{2017MNRAS.471.1390H}
{Hakobyan} A.~A.,  et~al., 2017, \mn@doi [\mnras] {10.1093/mnras/stx1608},
  \href {https://ui.adsabs.harvard.edu/abs/2017MNRAS.471.1390H} {471, 1390
  (H17)}

\bibitem[\protect\citeauthoryear{{Hakobyan}, {Barkhudaryan}, {Karapetyan},
  {Gevorgyan}, {Mamon}, {Kunth}, {Adibekyan}  \& {Turatto}}{{Hakobyan}
  et~al.}{2020}]{2020MNRAS.499.1424H}
{Hakobyan} A.~A.,  {Barkhudaryan} L.~V.,  {Karapetyan} A.~G.,  {Gevorgyan}
  M.~H.,  {Mamon} G.~A.,  {Kunth} D.,  {Adibekyan} V.,   {Turatto} M.,  2020,
  \mn@doi [\mnras] {10.1093/mnras/staa2940}, \href
  {https://ui.adsabs.harvard.edu/abs/2020MNRAS.499.1424H} {499, 1424 (H20)}

\bibitem[\protect\citeauthoryear{{Hakobyan}, {Karapetyan}, {Barkhudaryan},
  {Gevorgyan}  \& {Adibekyan}}{{Hakobyan} et~al.}{2021}]{2021MNRAS.505L..52H}
{Hakobyan} A.~A.,  {Karapetyan} A.~G.,  {Barkhudaryan} L.~V.,  {Gevorgyan}
  M.~H.,   {Adibekyan} V.,  2021, \mn@doi [\mnras] {10.1093/mnrasl/slab048},
  \href {https://ui.adsabs.harvard.edu/abs/2021MNRAS.505L..52H} {505, L52}

\bibitem[\protect\citeauthoryear{{Hillebrandt}, {Kromer}, {R{\"o}pke}  \&
  {Ruiter}}{{Hillebrandt} et~al.}{2013}]{2013FrPhy...8..116H}
{Hillebrandt} W.,  {Kromer} M.,  {R{\"o}pke} F.~K.,   {Ruiter} A.~J.,  2013,
  \mn@doi [Front. Phys.] {10.1007/s11467-013-0303-2}, \href
  {https://ui.adsabs.harvard.edu/abs/2013FrPhy...8..116H} {8, 116}

\bibitem[\protect\citeauthoryear{{Holwerda}, {Reynolds}, {Smith}  \&
  {Kraan-Korteweg}}{{Holwerda} et~al.}{2015}]{2015MNRAS.446.3768H}
{Holwerda} B.~W.,  {Reynolds} A.,  {Smith} M.,   {Kraan-Korteweg} R.~C.,  2015,
  \mn@doi [\mnras] {10.1093/mnras/stu2345}, \href
  {https://ui.adsabs.harvard.edu/abs/2015MNRAS.446.3768H} {446, 3768}

\bibitem[\protect\citeauthoryear{{Howell}}{{Howell}}{2001}]{2001ApJ...554L.193H}
{Howell} D.~A.,  2001, \mn@doi [\apjl] {10.1086/321702}, \href
  {https://ui.adsabs.harvard.edu/abs/2001ApJ...554L.193H} {554, L193}

\bibitem[\protect\citeauthoryear{{Howell} et~al.,}{{Howell}
  et~al.}{2009}]{2009ApJ...691..661H}
{Howell} D.~A.,  et~al., 2009, \mn@doi [\apj] {10.1088/0004-637X/691/1/661},
  \href {https://ui.adsabs.harvard.edu/abs/2009ApJ...691..661H} {691, 661}

\bibitem[\protect\citeauthoryear{{Juri{\'c}} et~al.,}{{Juri{\'c}}
  et~al.}{2008}]{2008ApJ...673..864J}
{Juri{\'c}} M.,  et~al., 2008, \mn@doi [\apj] {10.1086/523619}, \href
  {https://ui.adsabs.harvard.edu/abs/2008ApJ...673..864J} {673, 864}

\bibitem[\protect\citeauthoryear{{Kasparova}, {Katkov}, {Chilingarian},
  {Silchenko}, {Moiseev}  \& {Borisov}}{{Kasparova}
  et~al.}{2016}]{2016MNRAS.460L..89K}
{Kasparova} A.~V.,  {Katkov} I.~Y.,  {Chilingarian} I.~V.,  {Silchenko} O.~K.,
  {Moiseev} A.~V.,   {Borisov} S.~B.,  2016, \mn@doi [\mnras]
  {10.1093/mnrasl/slw083}, \href
  {https://ui.adsabs.harvard.edu/abs/2016MNRAS.460L..89K} {460, L89}

\bibitem[\protect\citeauthoryear{{Li} et~al.,}{{Li}
  et~al.}{2011}]{2011MNRAS.412.1441L}
{Li} W.,  et~al., 2011, \mn@doi [\mnras] {10.1111/j.1365-2966.2011.18160.x},
  \href {https://ui.adsabs.harvard.edu/abs/2011MNRAS.412.1441L} {412, 1441}

\bibitem[\protect\citeauthoryear{{Livio} \& {Mazzali}}{{Livio} \&
  {Mazzali}}{2018}]{2018PhR...736....1L}
{Livio} M.,  {Mazzali} P.,  2018, \mn@doi [\physrep]
  {10.1016/j.physrep.2018.02.002}, \href
  {https://ui.adsabs.harvard.edu/abs/2018PhR...736....1L} {736, 1}

\bibitem[\protect\citeauthoryear{{Maoz} \& {Mannucci}}{{Maoz} \&
  {Mannucci}}{2012}]{2012PASA...29..447M}
{Maoz} D.,  {Mannucci} F.,  2012, \mn@doi [\pasa] {10.1071/AS11052}, \href
  {https://ui.adsabs.harvard.edu/abs/2012PASA...29..447M} {29, 447}

\bibitem[\protect\citeauthoryear{{Onken} et~al.,}{{Onken}
  et~al.}{2019}]{2019PASA...36...33O}
{Onken} C.~A.,  et~al., 2019, \mn@doi [\pasa] {10.1017/pasa.2019.27}, \href
  {https://ui.adsabs.harvard.edu/abs/2019PASA...36...33O} {36, e033}

\bibitem[\protect\citeauthoryear{{Pan} et~al.,}{{Pan}
  et~al.}{2014}]{2014MNRAS.438.1391P}
{Pan} Y.~C.,  et~al., 2014, \mn@doi [\mnras] {10.1093/mnras/stt2287}, \href
  {https://ui.adsabs.harvard.edu/abs/2014MNRAS.438.1391P} {438, 1391}

\bibitem[\protect\citeauthoryear{{Panther}, {Seitenzahl}, {Ruiter}, {Crocker},
  {Lidman}, {Wang}, {Tucker}  \& {Groves}}{{Panther}
  et~al.}{2019}]{2019PASA...36...31P}
{Panther} F.~H.,  {Seitenzahl} I.~R.,  {Ruiter} A.~J.,  {Crocker} R.~M.,
  {Lidman} C.,  {Wang} E.~X.,  {Tucker} B.~E.,   {Groves} B.,  2019, \mn@doi
  [\pasa] {10.1017/pasa.2019.24}, \href
  {https://ui.adsabs.harvard.edu/abs/2019PASA...36...31P} {36, e031}

\bibitem[\protect\citeauthoryear{{Paturel} et~al.,}{{Paturel}
  et~al.}{1997}]{1997A&AS..124..109P}
{Paturel} G.,  et~al., 1997, \aaps, \href
  {http://adsabs.harvard.edu/abs/1997A%26AS..124..109P} {124}

\bibitem[\protect\citeauthoryear{{Phillips}}{{Phillips}}{1993}]{1993ApJ...413L.105P}
{Phillips} M.~M.,  1993, \mn@doi [\apjl] {10.1086/186970}, \href
  {https://ui.adsabs.harvard.edu/abs/1993ApJ...413L.105P} {413, L105}

\bibitem[\protect\citeauthoryear{{Piro}, {Thompson}  \& {Kochanek}}{{Piro}
  et~al.}{2014}]{2014MNRAS.438.3456P}
{Piro} A.~L.,  {Thompson} T.~A.,   {Kochanek} C.~S.,  2014, \mn@doi [\mnras]
  {10.1093/mnras/stt2451}, \href
  {https://ui.adsabs.harvard.edu/abs/2014MNRAS.438.3456P} {438, 3456}

\bibitem[\protect\citeauthoryear{{Raskin}, {Scannapieco}, {Rhoads}  \& {Della
  Valle}}{{Raskin} et~al.}{2009}]{2009ApJ...707...74R}
{Raskin} C.,  {Scannapieco} E.,  {Rhoads} J.,   {Della Valle} M.,  2009,
  \mn@doi [\apj] {10.1088/0004-637X/707/1/74}, \href
  {https://ui.adsabs.harvard.edu/abs/2009ApJ...707...74R} {707, 74}

\bibitem[\protect\citeauthoryear{{Roman} et~al.,}{{Roman}
  et~al.}{2018}]{2018A&A...615A..68R}
{Roman} M.,  et~al., 2018, \mn@doi [\aap] {10.1051/0004-6361/201731425}, \href
  {https://ui.adsabs.harvard.edu/abs/2018A&A...615A..68R} {615, A68}

\bibitem[\protect\citeauthoryear{{Ruiter} et~al.,}{{Ruiter}
  et~al.}{2013}]{2013MNRAS.429.1425R}
{Ruiter} A.~J.,  et~al., 2013, \mn@doi [\mnras] {10.1093/mnras/sts423}, \href
  {https://ui.adsabs.harvard.edu/abs/2013MNRAS.429.1425R} {429, 1425}

\bibitem[\protect\citeauthoryear{{Seth}, {Dalcanton}  \& {de Jong}}{{Seth}
  et~al.}{2005}]{2005AJ....130.1574S}
{Seth} A.~C.,  {Dalcanton} J.~J.,   {de Jong} R.~S.,  2005, \mn@doi [\aj]
  {10.1086/444620}, \href
  {https://ui.adsabs.harvard.edu/abs/2005AJ....130.1574S} {130, 1574}

\bibitem[\protect\citeauthoryear{{Shen}, {Toonen}  \& {Graur}}{{Shen}
  et~al.}{2017}]{2017ApJ...851L..50S}
{Shen} K.~J.,  {Toonen} S.,   {Graur} O.,  2017, \mn@doi [\apjl]
  {10.3847/2041-8213/aaa015}, \href
  {https://ui.adsabs.harvard.edu/abs/2017ApJ...851L..50S} {851, L50}

\bibitem[\protect\citeauthoryear{{Shen}, {Kasen}, {Miles}  \&
  {Townsley}}{{Shen} et~al.}{2018}]{2018ApJ...854...52S}
{Shen} K.~J.,  {Kasen} D.,  {Miles} B.~J.,   {Townsley} D.~M.,  2018, \mn@doi
  [\apj] {10.3847/1538-4357/aaa8de}, \href
  {https://ui.adsabs.harvard.edu/abs/2018ApJ...854...52S} {854, 52}

\bibitem[\protect\citeauthoryear{{Shen}, {Blondin}, {Kasen}, {Dessart},
  {Townsley}, {Boos}  \& {Hillier}}{{Shen} et~al.}{2021}]{2021ApJ...909L..18S}
{Shen} K.~J.,  {Blondin} S.,  {Kasen} D.,  {Dessart} L.,  {Townsley} D.~M.,
  {Boos} S.,   {Hillier} D.~J.,  2021, \mn@doi [\apjl]
  {10.3847/2041-8213/abe69b}, \href
  {https://ui.adsabs.harvard.edu/abs/2021ApJ...909L..18S} {909, L18}

\bibitem[\protect\citeauthoryear{{Sim}, {R{\"o}pke}, {Hillebrandt}, {Kromer},
  {Pakmor}, {Fink}, {Ruiter}  \& {Seitenzahl}}{{Sim}
  et~al.}{2010}]{2010ApJ...714L..52S}
{Sim} S.~A.,  {R{\"o}pke} F.~K.,  {Hillebrandt} W.,  {Kromer} M.,  {Pakmor} R.,
   {Fink} M.,  {Ruiter} A.~J.,   {Seitenzahl} I.~R.,  2010, \mn@doi [\apjl]
  {10.1088/2041-8205/714/1/L52}, \href
  {https://ui.adsabs.harvard.edu/abs/2010ApJ...714L..52S} {714, L52}

\bibitem[\protect\citeauthoryear{{Stritzinger}, {Leibundgut}, {Walch}  \&
  {Contardo}}{{Stritzinger} et~al.}{2006}]{2006A&A...450..241S}
{Stritzinger} M.,  {Leibundgut} B.,  {Walch} S.,   {Contardo} G.,  2006,
  \mn@doi [\aap] {10.1051/0004-6361:20053652}, \href
  {https://ui.adsabs.harvard.edu/abs/2006A&A...450..241S} {450, 241}

\bibitem[\protect\citeauthoryear{{Taubenberger}}{{Taubenberger}}{2017}]{2017hsn..book..317T}
{Taubenberger} S.,  2017, {in Alsabti~A.~W., Murdin~P., eds, The Extremes of
  Thermonuclear Supernovae, Handbook of Supernovae. Springer, p.~317}

\bibitem[\protect\citeauthoryear{{Yaron} \& {Gal-Yam}}{{Yaron} \&
  {Gal-Yam}}{2012}]{2012PASP..124..668Y}
{Yaron} O.,  {Gal-Yam} A.,  2012, \mn@doi [\pasp] {10.1086/666656}, \href
  {https://ui.adsabs.harvard.edu/abs/2012PASP..124..668Y} {124, 668}

\bibitem[\protect\citeauthoryear{{Yoachim} \& {Dalcanton}}{{Yoachim} \&
  {Dalcanton}}{2006}]{2006AJ....131..226Y}
{Yoachim} P.,  {Dalcanton} J.~J.,  2006, \mn@doi [\aj] {10.1086/497970}, \href
  {https://ui.adsabs.harvard.edu/abs/2006AJ....131..226Y} {131, 226}

\bibitem[\protect\citeauthoryear{{Yoachim} \& {Dalcanton}}{{Yoachim} \&
  {Dalcanton}}{2008}]{2008ApJ...683..707Y}
{Yoachim} P.,  {Dalcanton} J.~J.,  2008, \mn@doi [\apj] {10.1086/590246}, \href
  {https://ui.adsabs.harvard.edu/abs/2008ApJ...683..707Y} {683, 707}

\makeatother
\end{thebibliography}

%%%%%%%%%%%%%%%%% APPENDICES %%%%%%%%%%%%%%%%%%%%%

\appendix
\section{Online material}

Table~\ref{HOSTandSNIa} shows broadly binned morphological distribution
of SN~Ia subclasses in edge-on spiral galaxies.
Table~\ref{VVR25} compares the $|V|/Z_{25}$ distributions between different subclasses of SNe~Ia.
Table~\ref{ThickThindisc} compares the exponential scale heights of SN~Ia subclasses
with those of CC SNe, and thick and thin discs of galaxies.
Table~\ref{VV25dm15} shows the correlation test for the $|V|/Z_{25}$ versus $\Delta m_{15}$ parameters.

\begin{table}
  \centering
  \begin{minipage}{84mm}
  \caption{Broadly binned morphological distribution of SN~Ia subclasses in edge-on spiral host galaxies.}
  \tabcolsep 12pt
  \label{HOSTandSNIa}
    \begin{tabular}{lcccc}
    \hline
  \multicolumn{1}{l}{SN} & \multicolumn{1}{c}{S0/a--Sab} &   \multicolumn{1}{c}{Sb--Sc} &  \multicolumn{1}{c}{Scd--Sdm} & \multicolumn{1}{c}{All}\\
    \hline
    Normal & 46 & 81 & 17 & 144 \\
    91T & 5 & 18 & 7 & 30\\
    91bg & 13 & 10 & 0 & 23\\
    All & 67 & 108 & 22 & 197\\
  \hline
  \end{tabular}
  \end{minipage}
\end{table}
\begin{table}
  \centering
  \begin{minipage}{84mm}
  \caption{Comparison of the $|V|/Z_{25}$ distributions between different subclasses of SNe~Ia.}
  \tabcolsep 4.6pt
  \label{VVR25}
    \begin{tabular}{lcccccc}
    \hline
  Subsample~1 & \multicolumn{1}{c}{$N_{\rm SN}$} & versus & \multicolumn{1}{c}{Subsample~2} & \multicolumn{1}{c}{$N_{\rm SN}$} & \multicolumn{1}{c}{$P_{\rm KS}^{\rm MC}$}&\multicolumn{1}{c}{$P_{\rm AD}^{\rm MC}$}\\
  \hline
    Normal & 144 & versus & 91bg & 23 &  0.112 & \textbf{0.005}\\
    Normal & 144 & versus & 91T & 30 &  0.311 & 0.307\\
    91bg & 23 & versus & 91T & 30 &  \textbf{0.042} & \textbf{0.048}\\
  \hline
  \end{tabular}
  \small
  \parbox{\hsize}{\emph{Notes.} The explanations for the $P$-values are identical
                  to those in Table~\ref{UVPKSPAD}.}
  \end{minipage}
\end{table}
\begin{table}
  \centering
  \begin{minipage}{84mm}
  \caption{Comparison of exponential scale heights of SN~Ia subclasses with those of CC SNe,
           and thick and thin discs of edge-on galaxies.}
  \tabcolsep 7pt
  \label{ThickThindisc}
    \begin{tabular}{lccl}
    \hline
  \multicolumn{1}{l}{Disc} & \multicolumn{1}{l}{$N$} & \multicolumn{1}{c}{$\widetilde{H}$} & \multicolumn{1}{l}{Reference}\\
  \hline
  \multicolumn{4}{c}{Early-type galaxies}\\
    S0/a--Sc thin disc  & 122 & $0.02^{+0.01}_{-0.01}$ & \citet{2018AA...610A...5C} \\
    S0/a--Sab thin disc & 38 & $0.04^{+0.02}_{-0.01}$ & \citet{2018AA...610A...5C} \\
    \textbf{Normal (S0/a--Sab)} & 46 & $\textbf{0.07}^{+\textbf{0.03}}_{-\textbf{0.02}}$ & \textbf{This study}\\
    S0/a--Sc thick disc  & 122 & $0.11^{+0.02}_{-0.02}$ & \citet{2018AA...610A...5C} \\
    \textbf{91bg (S0/a--Sc)}  & 23 & $\textbf{0.14}^{+\textbf{0.08}}_{-\textbf{0.04}}$ &  \textbf{This study} \\
    \textbf{91bg (S0/a--Sab)}  & 13 & $\textbf{0.16}^{+\textbf{0.15}}_{-\textbf{0.06}}$ &  \textbf{This study} \\
    S0/a--Sab thick disc & 38 &  $0.17^{+0.07}_{-0.04}$ & \citet{2018AA...610A...5C} \\
    \multicolumn{4}{c}{Late-type galaxies}\\
    MW thin disc  &  & 0.02$\pm$0.01 & \citet{2008ApJ...673..864J}\\
    Sb--Sc thin disc & 84 & $0.02^{+0.01}_{-0.01}$ & \citet{2018AA...610A...5C}\\
    CC SNe  & 27 & $0.03^{+0.02}_{-0.01}$ & \citetalias{2017MNRAS.471.1390H}\\
    \textbf{91T (Sb--Sdm)} & 25 &  $\textbf{0.04}^{+\textbf{0.02}}_{-\textbf{0.01}}$ &  \textbf{This study}\\
    \textbf{Normal (Scd--Sdm)} & 17 &  $\textbf{0.05}^{+\textbf{0.04}}_{-\textbf{0.02}}$ & \textbf{This study}\\
    MW thick disc  & & 0.06$\pm$0.01 & \citet{2008ApJ...673..864J}\\
    \textbf{Normal (Sb--Sc)} & 81 & $\textbf{0.07}^{+\textbf{0.02}}_{-\textbf{0.01}}$ & \textbf{This study}\\
    Scd--Sd thick disc & 19 & $0.08^{+0.06}_{-0.03}$ & \citet{2018AA...610A...5C} \\
    Sb--Sc thick disc & 84 & $0.08^{+0.02}_{-0.02}$ & \citet{2018AA...610A...5C} \\
    \textbf{91bg (Sb--Sc)}  & 10 & $\textbf{0.12}^{+\textbf{0.13}}_{-\textbf{0.05}}$ &  \textbf{This study} \\
  \hline
  \end{tabular}
  \small
  \parbox{\hsize}{\emph{Notes.} $\widetilde{H}_{\rm SN}=\langle|V|/R_{25}\rangle$.
                  Morphological classification of galaxies from \citet{2018AA...610A...5C}
                  is available via the HyperLeda and/or NED.
                  The $\widetilde{H}$ values are displayed in ascending order.}
  \end{minipage}
\end{table}
\begin{table}
  \centering
  \begin{minipage}{84mm}
  \caption{The correlation test for the $|V|/Z_{25}$ versus $\Delta m_{15}$ parameters.}
  \tabcolsep 5.8pt
  \label{VV25dm15}
    \begin{tabular}{lcccccc}
    \hline
  SN & \multicolumn{1}{c}{$N_{\rm SN}$} & \multicolumn{1}{c}{$\langle|V|/Z_{25}\rangle$} & versus & \multicolumn{1}{c}{$\langle\Delta m_{15}\rangle$} & \multicolumn{1}{c}{$r_{\rm s}$} & \multicolumn{1}{c}{$P_{\rm s}^{\rm MC}$} \\
  \hline
    All & 69 & $0.24^{+0.07}_{-0.05}$ & versus & 1.21$\pm$0.32 &  0.058 & 0.633 \\
    All$\dag$ & 36 & $0.41^{+0.18}_{-0.11}$ & versus & 1.18$\pm$0.29 &  0.396 & \textbf{0.016} \\
  \hline
  \end{tabular}
  \small
  \parbox{\hsize}{\emph{Notes.} For the explanations of the parameters see Table~\ref{VR25dm15}.
                  The subsample marked with $\dag$ symbol corresponds to the dust-truncated SNe
                  with $|V|/Z_{25} \geq 0.123$.
                  For our host galaxy sample, the truncation is obtained by considering that
                  $\langle Z_{25}/R_{25}\rangle \approx0.325$.}
  \end{minipage}
\end{table}

A portion of the database underlying the study
is shown in Table~\ref{databasetab} for guidance regarding its content and format.
The entire table is available in electronic format as an CSV file.

\begin{sidewaystable}
  \caption{The database of 197 SNe~Ia and their 196 host galaxies.
           The first ten entries are displayed.
           The full table can be found in the article's online version.}
  \label{databasetab}
  \begin{tabular}{lccccccccccc}
    \hline
    \multicolumn{1}{c}{SN} & \multicolumn{1}{c}{Subclass} & \multicolumn{1}{c}{Source} & \multicolumn{1}{c}{$U$} & \multicolumn{1}{c}{$V$} & \multicolumn{1}{c}{$\Delta m_{15}$} & \multicolumn{1}{c}{Source} & \multicolumn{1}{c}{Host} & \multicolumn{1}{c}{Dist.}  & \multicolumn{1}{c}{Morph.} & \multicolumn{1}{c}{$R_{25}$} & \multicolumn{1}{c}{$Z_{25}$}\\
     & & \multicolumn{1}{c}{Bibcode} & \multicolumn{1}{c}{arcsec} & \multicolumn{1}{c}{arcsec} & \multicolumn{1}{c}{mag} & \multicolumn{1}{c}{Bibcode}  & & \multicolumn{1}{c}{Mpc} & & \multicolumn{1}{c}{arcsec} & \multicolumn{1}{c}{arcsec} \\
    \hline
    1959C & norm & 1993AJ....106.2383B & 8.340 & 0.066 & $0.74\pm0.07$ & 2017ApJ...835...64G & MCG+01-34-005 & 42.563 & Sc & 44.826 & 11.715\\
    1962J & norm & 1993AJ....106.2383B & 40.100 & 13.776 & $1.01\pm0.07$ & 2017ApJ...835...64G & NGC6835 & 20.201 & Sab & 89.277 & 29.850\\
    1984A & norm & 2009ApJ...699L.139W & 25.132 & 13.913 & $1.21\pm0.10$ & 2005ApJ...623.1011B &  NGC4419 & 16.095 & Sa & 105.893 & 44.721\\
    1990G & norm & 2012MNRAS.425.1789S & 35.170 & 5.912 & - & - & IC2735 & 153.410 & Sab & 40.386 & 14.789\\
    1991bd & norm & 2012MNRAS.425.1789S & 35.801 & 6.736 & - & - & UGC02936 & 51.771 & Sc & 117.645 & 22.623\\
    1991bf & norm & 2012MNRAS.425.1789S & 25.416 & 1.118 & - & - & ESO471-030 & 120.768 & Sa & 39.665 & 17.126\\
    1991K & 91T & 2012MNRAS.425.1789S & 18.898 & 7.251 & - & - & NGC2851 & 70.776 & S0/a & 62.163 & 30.627\\
    1992ag & norm & 1992IAUC.5555....1M & 2.849 & 2.067 & $1.19\pm0.10$ & 1996AJ....112.2408H & ESO508-067 & 104.045 & Sb? & 33.522 & 10.278\\
    1993ah & norm & 1993IAUC.5897....1B & 7.661 & 0.914 & $1.30\pm0.10$ & 1996AJ....112.2408H & ESO471-027 & 120.357 & Sab & 36.384 & 13.386\\
    1993L & norm & 1993IAUC.5782....1D & 23.490 & 4.488 & $1.47\pm0.07$ & 2005ApJ...624..532R & IC5270 & 23.387 & Sbc & 71.856 & 20.562\\
    \hline
  \end{tabular}
  \parbox{\hsize}{\emph{Notes.} We more precisely identify the morphologies of edge-on galaxies
                  by assessing the size of the bulge relative to the disc (see \citetalias{2017MNRAS.471.1390H}).}
\end{sidewaystable}

\label{lastpage}

\end{document}